\documentclass[12pt]{article}
\usepackage{amsmath}
\usepackage{ulem}
\usepackage{graphicx}
\usepackage{amssymb}

        \newdimen\eqskip
        \newdimen\txtskip
        \baselineskip=\txtskip
        \newdimen\mysep                
        \newdimen\hmysep
\begin{document}

  \newcommand{\ccaption}[2]{
    \begin{center}
    \parbox{0.9\textwidth}{
      \caption[#1]{\small{{#2}}}
      }
    \end{center}
    }
\newcommand{\BS}{\bigskip}
\def\plaga{\cite{Plaga:2008gq}}
\def\casadio{\cite{Casadio:2001wh}}
\def\GM{\cite{Giddings:2008gr}}
\def    \be             {\begin{equation}}
\def    \ee             {\end{equation}}
\def    \ba             {\begin{eqnarray}}
\def    \ea             {\end{eqnarray}}
\def    \nn             {\nonumber}
\def    \=              {\;=\;}
\def    \ret            {\\[\eqskip]}
\def    \ie             {{\em i.e.\/} }
\def    \eg             {{\em e.g.\/} }
\def \lsim{\lesssim}
\def \gsim{\gtrsim}
\def    \ev            {\mbox{$\mathrm{eV}$}}
\def    \kev            {\mbox{$\mathrm{keV}$}}
\def    \mev            {\mbox{$\mathrm{MeV}$}}
\def    \gev            {\mbox{$\mathrm{GeV}$}}
\def	\tev		{\mbox{$\mathrm{TeV}$}}
\def    \gr            {\mbox{$\mathrm{gr}$}}
\def    \fm            {\mbox{$\mathrm{fm}$}}
\def    \cm            {\mbox{$\mathrm{cm}$}}
\def    \km            {\mbox{$\mathrm{km}$}}
\def    \isec            {\mbox{$\mathrm{s}^{-1}$}}
\def    \yr            {\mbox{$\mathrm{yr}$}}
\def\rem           {\ifmmode R_{\mathrm{EM}} \else $R_{\mathrm{EM}}$ \fi}
\def\remsq           {\ifmmode R^2_{\mathrm{EM}} \else $R^2_{\mathrm{EM}}$ \fi}
\def\remf         {\ifmmode R_{\mathrm{EM,4}} \else $R_{\mathrm{EM,4}}$ \fi}
\def\Rs {R}
\def\ktd {{\tilde k}_D}
\def\kd {k_D}
\def\khd{{\hat k}_D}
\def\arad{a}
\def\rson{R_B}
\def\hf{{1\over 2}}
\begin{titlepage}
\nopagebreak
{\begin{flushright}{
 \begin{minipage}{5cm}
 {
   CERN-PH-TH/2008-184}
\end{minipage}}\end{flushright}}
\vfill
\begin{center}
{\bf\sc\LARGE Comments on claimed risk 
  \\[0.5cm]
from metastable black holes}
\end{center}
\vfill
\begin{center}
{\large 
Steven B. Giddings$^{a,b,}$\footnote{giddings@physics.ucsb.edu}
 and Michelangelo
 L.~Mangano$^{b,}$\footnote{michelangelo.mangano@cern.ch}
}
\end{center}

\vskip 0.5cm

{\it  $^{a}$ Department of Physics, 
University of California, Santa Barbara, CA 93106}

{\it $^{b}$ PH-TH, CERN, Geneva, Switzerland
}
\vfill
\begin{abstract}
In a recent note, {\tt arXiv:0808.1415}, it was argued that a hypothetical metastable black hole scenario could pose collider risk not excluded by our previous study.  We comment on inconsistency of this proposed scenario.
\end{abstract}
CERN-PH-TH/2008-184\hfill \\
\today \hfill
\vfill
\end{titlepage}
\newpage

The recent paper \cite{Plaga:2008gq} claims to produce a scenario for
LHC risk that is not excluded by our previous
work\cite{Giddings:2008gr}, or by the paper
\cite{Koch:2008qq} supporting the same conclusions.
This proposal is based on an idea,
extrapolating  claims in \cite{Casadio:2001wh}, that Hawking radiation
is suppressed until a certain mass threshold, which
\cite{Plaga:2008gq} proposes occurs for black hole radii comparable to
the scale where higher-dimensional gravity matches onto
four-dimensional gravity.  Ref.~\cite{Plaga:2008gq} claims that once
the Hawking radiation turns on, the power output would be at  dangerous
levels like $10^{16}$~W, and that moreover such a scenario is not
excluded by astrophysical constraints such as those derived
in~\cite{Giddings:2008gr} from white dwarfs and neutron stars.

To assess this scenario, let us begin by observing that a universal
relation for the energy output of Hawking radiation  
is of the form  
\be \label{eq:HR}
\frac{dE}{dt} =  \xi \frac{(D-3)^4}{3840 \pi} \; \frac{1}{R^2} \sim
(D-3)^4 \times 10^{-4} \; \frac{1}{R^2} \; ,
\ee
where $D$ is the spacetime dimension, $R$ is the Schwarzschild radius, and $\xi$ is an ${\cal O}(1)$ gray-body factor parameterizing the deviations from a precise blackbody spectrum.  This general formula follows from two facts:  the Stefan-Boltzmann law, and the formula for the Hawking temperature of the black hole,
\be
\label{eq:TvsR}
T \; = \; \frac{D-3}{4\pi \; R}  \; .
\ee
Moreover, this formula agrees with eq.~(2) of \cite{Plaga:2008gq}.

The proposal of \cite{Plaga:2008gq} is that Hawking radiation is
suppressed compared to this usual Hawking result (hence the black hole
is ``metastable") until the black hole reaches the radius scale $R\sim
L$, where $L$ is comparable to the scale of transition to
four-dimensional behavior.  (This could be anywhere in the range
between the parameters $R_D$ and $R_C$ introduced in
\cite{Giddings:2008gr}.)  When it reaches this scale, the usual
Hawking radiation is then claimed to switch on.
Ref.~\cite{Plaga:2008gq} considers in particular scales near $L\sim
10^{-5} \cm$, where one is claimed to find the large power output
stated above.

However, using these parameters one readily finds from the formula (\ref{eq:HR}) a negligible power output of size 
\be\label{micwat}
\frac{dE}{dt}\sim .1 \mu W\ ,
\ee
differing by a factor of $10^{23}$ from the claim of \plaga.

Where did \cite{Plaga:2008gq} go wrong?  The answer is in the
inconsistent application of formula (2) of that paper.  In the type of
warped scenario that \cite{Plaga:2008gq} considers, the black hole
would evolve up to a radius $R\sim R_D$ via higher-dimensional
evolution, and then would experience a large mass gain in
transitioning to a slightly higher radius $R\sim R_C$, as \plaga\ 
acknowledges.  Throughout
this region, in the usual Hawking scenario, the temperature formula
(\ref{eq:TvsR}) should hold. Thus if the black hole radiance is suppressed
compared to this, as the author of \plaga\ proposes, 
it can't exceed a value of size~(\ref{micwat}).
However, \cite{Plaga:2008gq} then applies the formula (\ref{eq:HR}) written
in terms of the mass using the four-dimensional relationship between
radius and mass, but does this in a region where the four-dimensional
relation between radius and mass is clearly wrong.  Indeed, the
four-dimensional Schwarzschild radius corresponding to the mass
range considered by \cite{Plaga:2008gq}, $1 \lsim
M(\mathrm{kg})\lsim 10^5$, lies in the range $10^{-25}-10^{-20}\cm$,
far below  the claimed $\sim 10^{-5}\cm$!  It is this inconsistency
that produces the claimed large power output, which, if correct, would
represent an enormous {\it enhancement} of the black hole radiance, in
contradiction to the stated assumptions of the scenario.

In addition to this basic inconsistency, there are other arguments
against such a proposal; we note these here, and defer further
explanation for future comment.  First, in such a ``suppressed
Hawking radiation" scenario, the arguments of \cite{Giddings:2008gr}
tell us that one can in fact not establish Eddington-limited accretion
in a white dwarf (see in particular eq.(B.13)
in~\cite{Giddings:2008gr}); 
so, even ignoring the inconsistency we reported above, 
the bounds of that paper would apply even to the suggested scenario
of~\cite{Plaga:2008gq}. 
Second, the underlying basis, namely a
serious difference between the microcanonical picture and the usual
Hawking calculation, appears implausible in the large black hole
regime~\cite{Plaga:2008gq} considers. 

Finally, we note that \cite{Plaga:2008gq} has both misquoted our
paper \GM, and selectively quoted from the available literature.  The
correct statement, misquoted in footnote 3 of \cite{Plaga:2008gq},
states: ``...at each point where we have encountered an uncertainty,
we have replaced it by a conservative or ``worst case" assumption."
Moreover, \cite{Plaga:2008gq} cites Unruh and Sch\"utzhold's
work\cite{Unruh:2004zk} raising questions about Hawking radiation,
without providing the more recent citation to Unruh's work,
\cite{Unruh}, that was given in \cite{Giddings:2008gr} and reflects
more up-to-date comments by Unruh on the support of his work for
Hawking radiation.

 We conclude that the conclusions of~\cite{Giddings:2008gr} on this
 subject, as
stated there and as 
referred to in the LHC safety assessment report~\cite{Ellis:2008hg},
remain robust.

\end{document}